\documentclass{article}

\usepackage{arxiv}

\usepackage[utf8]{inputenc} 
\usepackage[T1]{fontenc}    
\usepackage[hidelinks]{hyperref}       
\usepackage{url}            
\usepackage{booktabs}       
\usepackage{amsfonts,amsmath}       
\usepackage{nicefrac}       
\usepackage{microtype}      
\usepackage{lipsum}
\usepackage{graphicx}
\usepackage{bm}
\graphicspath{ {./images/} }

\newcommand\blfootnote[1]{%
  \begingroup
  \renewcommand\thefootnote{}\footnote{#1}%
  \addtocounter{footnote}{-1}%
  \endgroup
}

\title{The role of physical models in the validation \\
and calibration of numerical models \\
– The example of the Lillebælt Bridge}

\author{
 Paula Apollonia Wunderlich$^{\star}$ \\
  Bauhaus-Universität Weimar\\
  Weimar, Germany \\
  \And
 Gledson Rodrigo Tondo \\
  Bauhaus-Universität Weimar\\
  Weimar, Germany \\
  \And
 Guido Morgenthal \\
  Bauhaus-Universität Weimar\\
  Weimar, Germany \\
}

\begin{document}
\maketitle
\begin{abstract}
With the rapid advancement of computer technologies enabling fast calculations of complex structures, numerical methods have become a central tool in engineering sciences, while physical models have increasingly receded into the background. Nevertheless, owing to their clarity and comprehensibility, these former engineering tools remain of great value and their use can still be highly relevant today. At the example of the scale model of the Lillebælt Bridge – developed by the Copenhagen engineers Christen Ostenfeld and Wriborg Jønson and given for research purposes to the Bauhaus-Universität Weimar – this paper illustrates how physical models can still serve as useful instruments in research and teaching. By applying operational modal analysis, the natural frequencies and damping ratios of the bridge model are experimentally determined, which in turn can serve as reference data for the calibration and validation of numerical models.
\end{abstract}


\section{Introduction}\blfootnote{IABSE Symposium Copenhagen 2026 - Bridging Advanced Technologies - Structural Innovation\\ \hspace*{5mm} Copenhagen, DK, 2026}

Physical models serve to make complex phenomena more comprehensible by making them tangible. In the context of civil engineering, they are primarily employed to elucidate load bearing mechanisms and the dynamic behaviour of structures. A significant milestone in structural model analysis was set by the mathematician Leonhard Euler in the 18\textsuperscript{th} century, when he addressed the fundamental question of how results obtained from model tests can be scaled to the actual structure \cite{Addis.2021}. Driven by the inventive spirit and audacity of engineers and architects such as Fritz Leonhardt, Frei Otto and Heinz Hossdorf, the structural analysis through physical models found systematic application in the 20\textsuperscript{th} century \cite{Muller.1971, Weber.2011}.

Scale models played an important role in the planning phase of the Lillebælt Bridge. The Copenhagen engineers Christen Ostenfeld and Wriborg Jønson, who were responsible for designing the bridge, developed a three-dimensional scale model of the structure to verify the load-bearing capacity and dynamic behaviour of the final design. Today, this model is preserved at the Bauhaus-Universität Weimar, where it continues to serve research purposes.

Even though numerical models have become the central tool of engineering sciences today and detailed scale models are hardly built anymore to assess the static and dynamic behaviour of structures, the value of historical experimental models should not be underestimated. They continue to contribute to an illustrative understanding of complex structural behaviour and can therefore be used to great effect in research and teaching. In this context, it is worth mentioning the research project ``Last Witnesses’ Measurement Models in Civil Engineering – Scientific Significance and Preservation'', funded as part of the priority program 2255 ``Construction as Cultural Heritage'' of the German Research Foundation, as it aims to preserve the significant heritage of physical models \cite{Schmid.2024}. 

This paper presents a dynamic analysis of the Lillebælt scale model, the results of which can then serve as reference data for validating and calibrating numerical models.

\section{The physical model of the Lillebælt Bridge} \label{sec:2}
\subsection{Motivation for the scale model}
Built between 1965 and 1970, the Lillebælt Bridge was Denmark’s first suspension bridge. With a main span of 600~m and two side spans of 240~m each, it connects the Jutland peninsula with the island of Funen \cite{Ostenfeld.1970}.

At the time of its completion, the Lillebælt Bridge was one of the first suspension bridges whose superstructure consisted of flat box girders. In response to the collapse of the Tacoma Narrows Bridge in 1940 as a result of torsional flutter, massive, heavy truss girders were predominantly used in the subsequent years to avoid aerodynamic instabilities \cite{Scott.2001}. Such a solution had initially been planned for the Lillebælt Bridge as well. However, Ostenfeld and Jønson chose to replace the original truss design with a lighter flat box girder. This not only offered significant reductions in weight and material consumption but also provided greater torsional stiffness \cite{Leonhardt.1984}. 

Since the use of flat box girders in suspension bridges had been scarcely explored at that time, the engineers at Chr.~Ostenfeld~\&~W.~Jønson Consulting Engineers carried out extensive investigations using physical scale models. Model tests provided valuable insights into the static and dynamic behaviour of the structure and played a decisive role in strengthening confidence in the new design and proving its performance.

\subsection{Model tests}
During the design phase, the engineers at Chr.~Ostenfeld~\&~W.~Jønson Consulting Engineers developed a two-dimensional (2D) as well as a three-dimensional (3D) model of the Lillebælt Bridge, representing the real construction on a scale of 1:200.

The 3D model was used to determine the natural frequencies for comparison with the calculated values. In addition, the spatial model served to investigate relevant deformations under traffic loads as well as under horizontal loads, in particular mean wind loads. Based on the 3D model, influence lines for deflections were obtained in order to demonstrate the load-distribution capacity of the torsionally rigid section as well as the equalising effect upon the cables resulting from the torsional rigidity of the towers. The 2D model, on the other hand, was used to study the effect of various boundary conditions – such as couplings between the stiffening girder and the main cables, the continuity of the stiffening girder and the use of inclined hangers – in order to establish the final design of the static system \cite{Ostenfeld.1970}.

\begin{figure}[h]
\centering
\includegraphics[width=10cm]{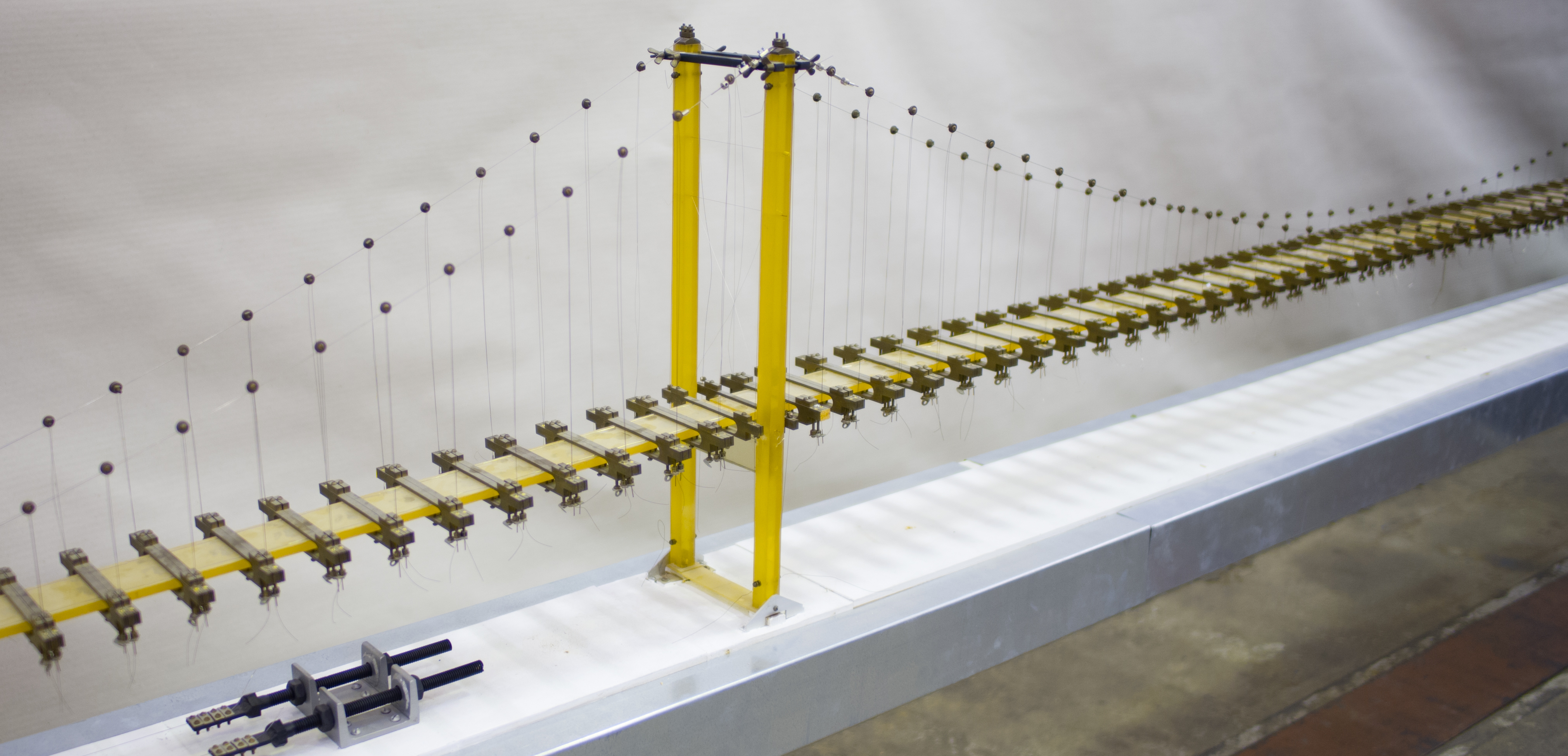}
\caption{1:200 scale model of the Lillebælt Bridge (Photo by Guido Morgenthal).}
\label{fig:01}
\end{figure}
The 3D model, shown in Figure \ref{fig:01}, has been well preserved and stands as an impressive testament to the remarkable engineering technologies of its time. Initially kept at COWI, it was later provided to the Bauhaus-Universität Weimar for research and teaching purposes. This became possible through Guido Morgenthal, who had worked at COWI before being appointed to a professorship at the Bauhaus-Universität.

\subsection{Model description}
The scaled 3D model of the Lillebælt Bridge has an overall length of 540~cm, comprising a 300~cm main span and two side spans of 120~cm each. The pylons of the model reach a height of 60~cm, while the end piers measure 18~cm.

\begin{figure}[h]
\centering
\includegraphics[width=8cm]{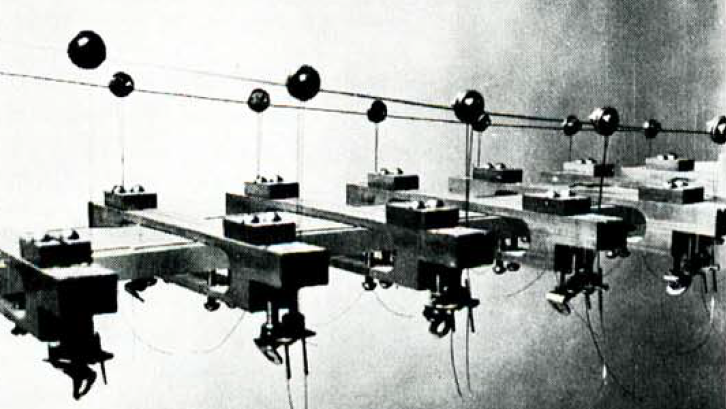}
\caption{Detail of the Lillebælt scale model showing the model superstructure \cite{Ostenfeld.1970}.}
\label{fig:02}
\end{figure}
For accurate investigation of the bridge’s dynamic behaviour, precise scaling of mass and stiffness is essential to adequately represent the real properties of the structure. This also governs the design of the model superstructure, which consists of two components: brass elements that provide the required mass and an acrylic glass plate that provides the necessary stiffness (see Figure \ref{fig:02}). An interruption in the acrylic glass plate at the location of the pylons ensures the main field to be decoupled from the side fields. Vertical hangers made of stainless-steel wire connect the superstructure to the main suspension cables via tensile forces. The main cables, also made of stainless-steel wire, are guided over the pylons and lie there in a cable saddle. The pylons transfer vertical force components of the main cables into the ground, while the horizontal components are absorbed by anchoring the cables at the end of the model bridge.

\section{Dynamic investigations on the physical model}
Data obtained from vibration tests provide valuable insights into the structure’s local and global behaviour and can be acquired in a non-destructive manner. They are therefore particularly well suited as reference data for numerical model updating. When experimental data is thereby derived from a scale model, the results can be transferred to full scale using similitude laws \cite{Hossdorf.1971}.

The following section presents the experimental determination of the natural frequencies and damping ratios of the Lillebælt scale model.




\subsection{Identification of the natural frequencies}
The focus is on determining the natural frequencies corresponding to the first three bending modes and to the first torsional mode of the bridge model. These are identified using output-only modal analysis, whereby the structural response to an external excitation of the physical model is recorded in terms of accelerations in global z-direction and angular velocities about the global x-axis. The recorded sensor data is then transformed into frequency domain using fast Fourier transform (FFT) to extract the eigenfrequencies.

\subsubsection{Excitation}
The system is set into vibration by an impulse excitation. This type of excitation covers a broad frequency band, ensuring that all modes of interest can be sufficiently excited. 

\begin{figure}[h]
\centering
\includegraphics[width=6cm]{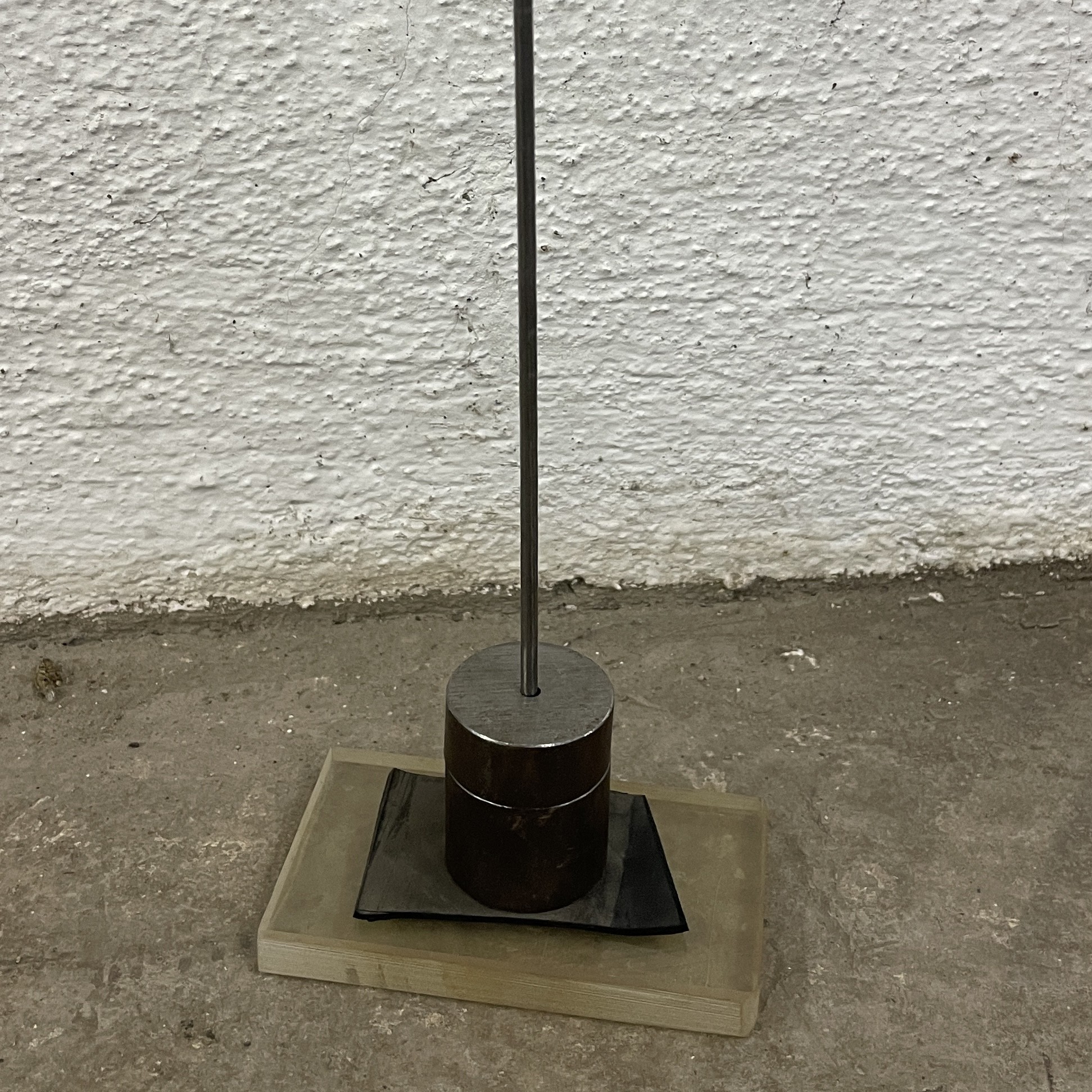}
\caption{Device for impulse excitation.}
\label{fig:03}
\end{figure}
For the impulse excitation, the device shown in Figure \ref{fig:03}, which was already used in the former dynamic tests on the Lillebælt scale model, is placed on the model superstructure. It consists of a single mass of $m$~=~192~g that is lifted along the rod and then dropped. The weight of the entire excitation device is 263~g. To capture the natural frequencies as accurately as possible, excitation is applied close to locations where maximum deflections in the mode shapes are expected. 

\subsubsection{Sensor selection and configuration}
\begin{figure}[h]
\centering
\includegraphics[width=6cm]{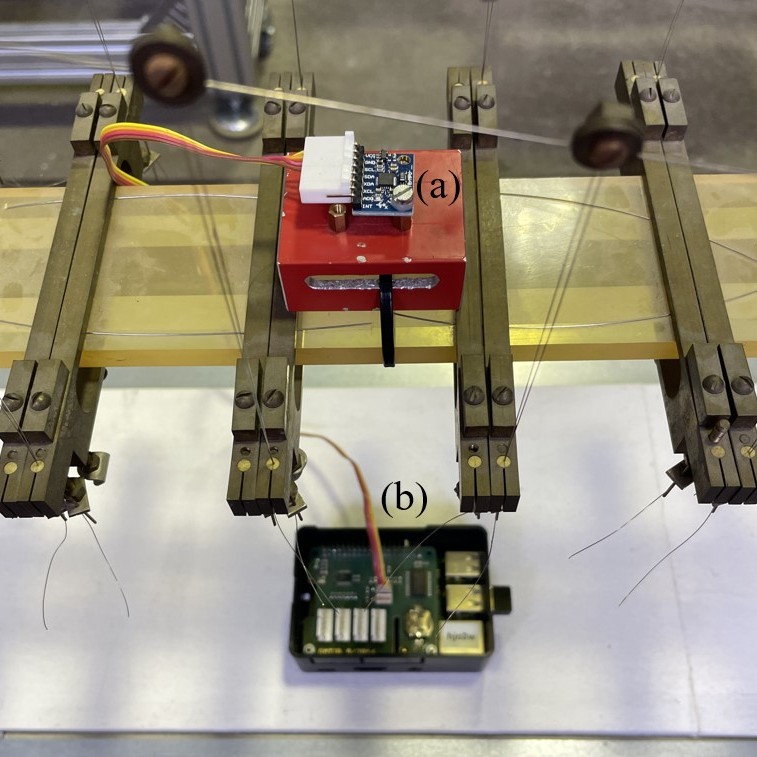}
\caption{Sensor mounting (a) on the model superstructure together with the Raspberry Pi (b).}
\label{fig:04}
\end{figure}

MPU-6050 sensors, that integrate a three-axis accelerometer with a three-axis gyroscope \cite{InvenSenseInc..2013}, are employed in the experiment. The MEMS-based sensors, controlled by Raspberry Pi microcomputers \cite{Morgenthal.2019}, are mounted on a metal frame and centrally positioned on top of the superstructure (see Figure \ref{fig:04}).

The MPU-6050 sensors have an acceleration measurement range of ±4~g and an angular velocity measurement range of ±1000~°/s \cite{InvenSenseInc..2013}, which adequately covers the expected accelerations and angular velocities.

To ensure that the structural response is captured without information loss, the sensor sampling rate must be sufficiently high. According to the Nyquist-Shannon sampling theorem, this condition is satisfied if the sampling frequency is at least twice the highest frequency present in the signal \cite{Kuttner.2019}. The expected frequency range for the experiment can be estimated based on previous tests on the Lillebælt scale model. The highest natural frequency of interest in this study is the first torsional eigenfrequency, which is anticipated to be approximately 8~Hz. Following the sampling theorem, this requires a minimum sampling rate of 16~Hz. In general, a higher sampling rate improves measurement accuracy. However, the sampling rate should not be set arbitrarily high in order to avoid an unnecessarily large amount of sensor data. Therefore, a sampling frequency of 200~Hz was chosen.

\begin{figure}[h]
\centering
\includegraphics[width=8cm]{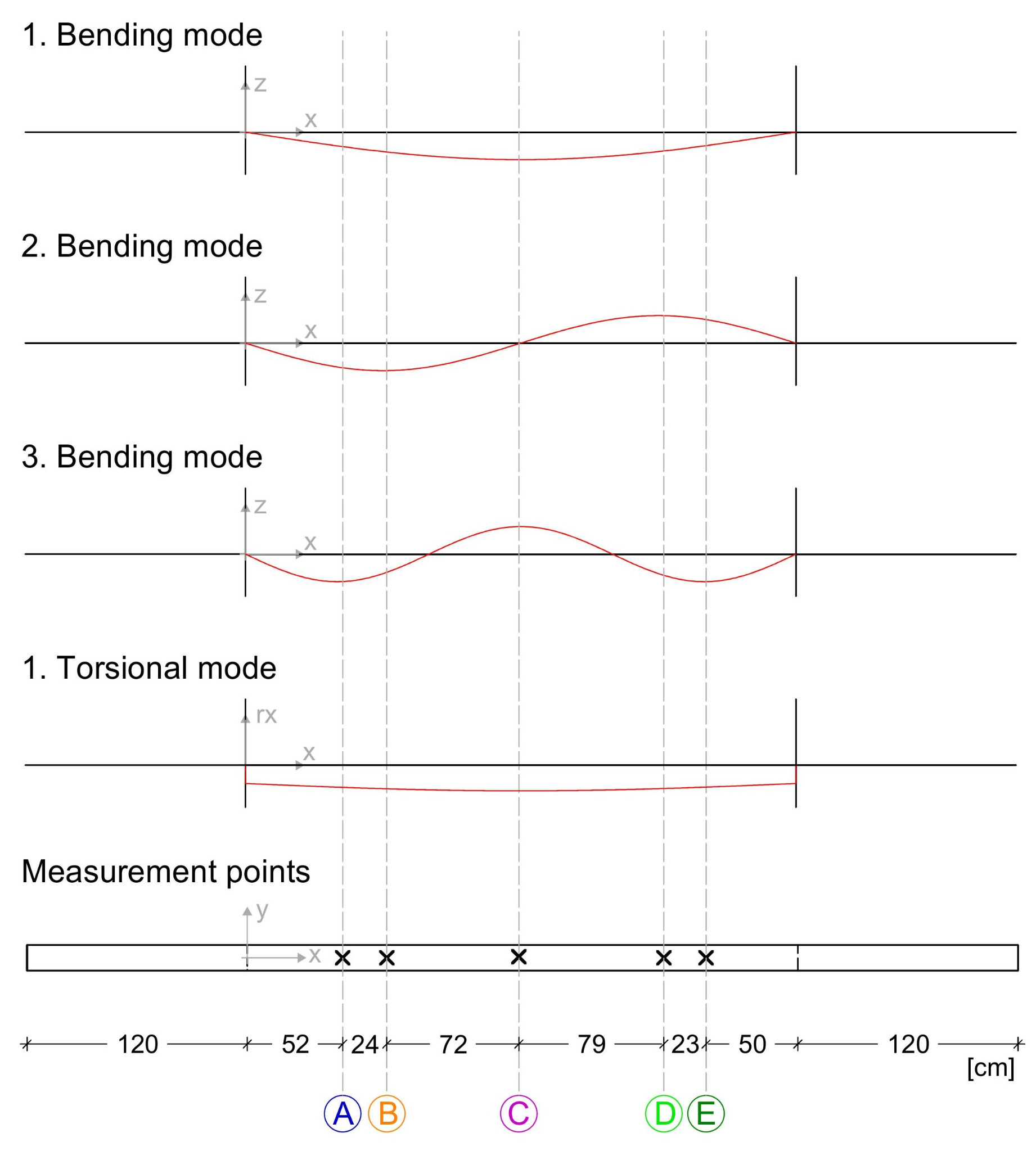}
\caption{Mode shapes of the scale model along with measurement points.}
\label{fig:05}
\end{figure}

To capture the desired mode shapes and the corresponding natural frequencies as accurately as possible, sensors should be positioned at locations where maximum deflections in the dynamic structural response are expected. Figure \ref{fig:05} illustrates the first three bending modes and the first torsional mode of the model main span, that were derived from an uncalibrated finite element (FE) model, together with the selected measurement points. Due to the design of the model superstructure, some of the sensors cannot be placed directly at the point of maximum deflection, but only in their immediate vicinity. Measurement point C covers the maximum of the first bending mode as well as one of the maxima of the third bending mode. The two remaining maxima of the third bending mode are captured by measurement points A and E. The maxima of the second bending mode are covered by measurement points B and D. The first torsional mode is well captured by all measurement points. For the eigenfrequency identification tests, three sensors are arranged in two setups along the superstructure, covering all measurement points.

\subsubsection{Results}
The recorded acceleration time series in global z-direction and the angular velocity time series about global x-axis are transformed into frequency domain by means of an FFT. The peaks in the resulting frequency spectra indicate the natural frequencies.

\begin{figure}[h]
\centering
\includegraphics[width=8cm]{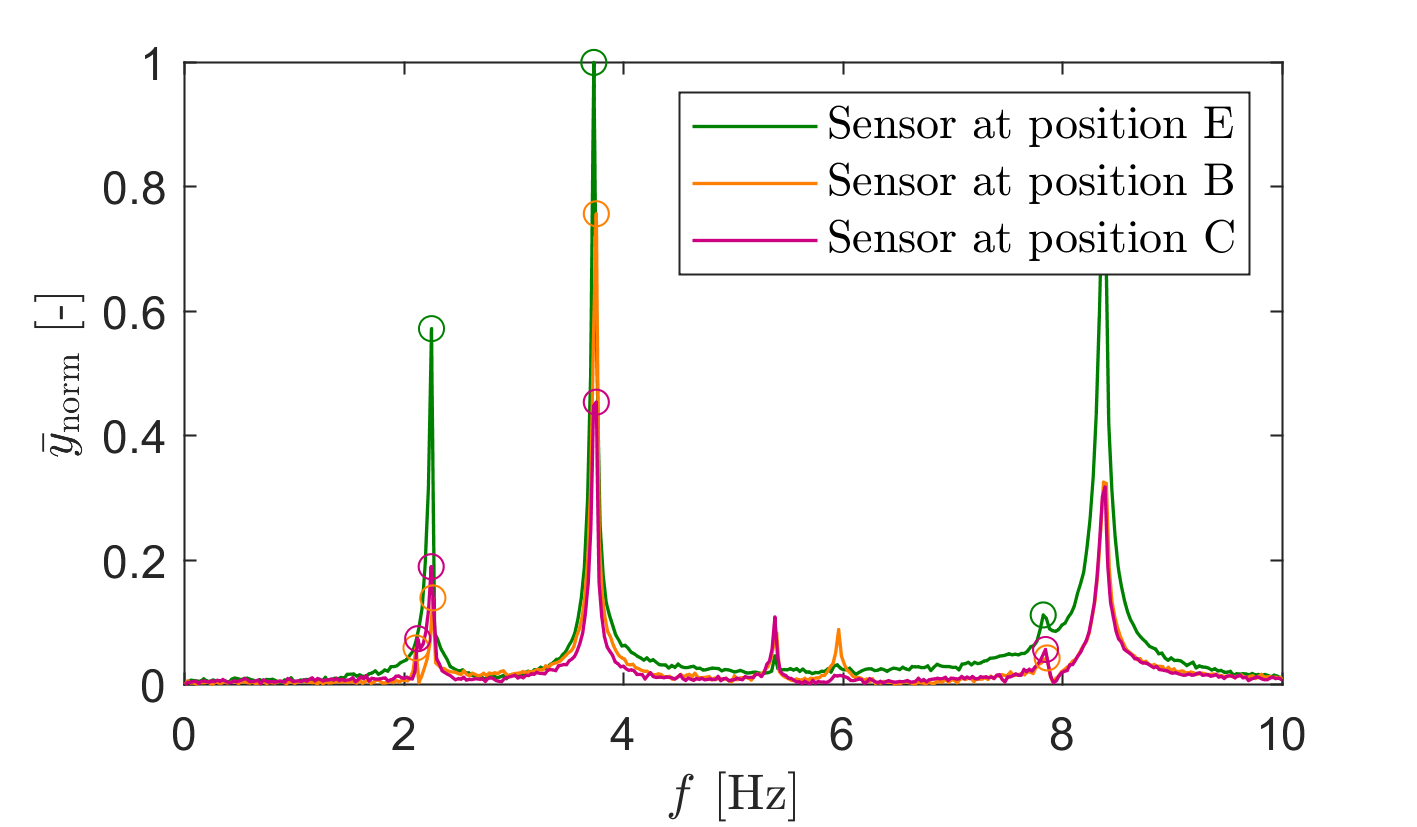}
\includegraphics[width=8cm]{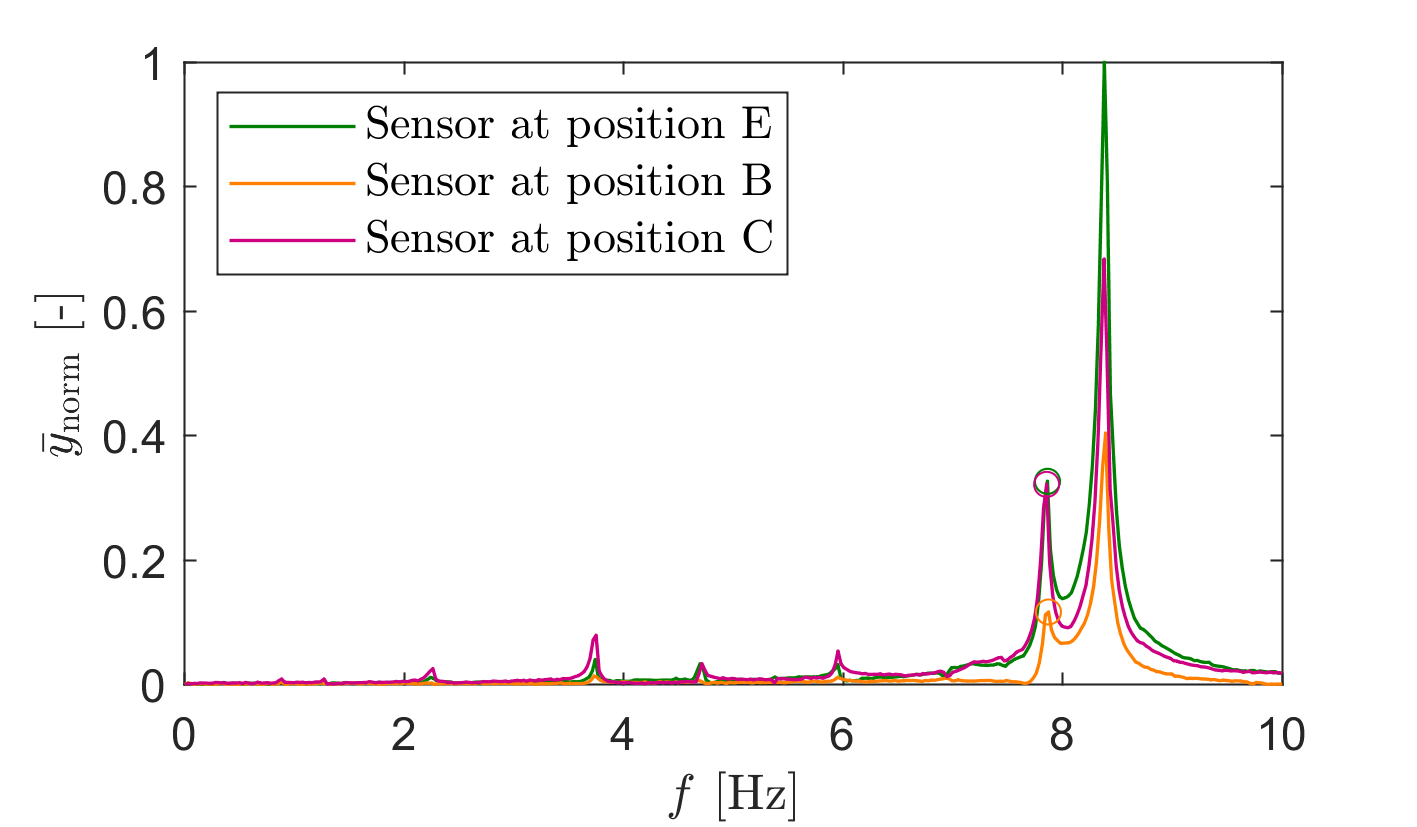}
\caption{Normalized frequency spectrum of the acceleration response in z, with peaks indicating the first three bending mode frequencies (left) and normalized frequency spectrum of the angular velocity response around x, with peaks indicating the first torsional mode frequency (right).}
\label{fig:0607}
\end{figure}

Figure \ref{fig:0607} presents the frequency spectrum obtained from accelerations and angular velocities, recorded in a representative measurement. The peaks in the spectra are normalized to the maximum peak occurring within the displayed frequency range. Additional peaks in the spectrum that are not marked correspond to other mode shapes that are not targeted in this work.

In total, six measurements were carried out. In each measurement, the impulse excitation was applied at a different location along the main span.
The eigenfrequencies identified from all measurements are compiled, and their mean value is calculated. The results of the natural frequencies of the Lillebælt scale model, along with the corresponding standard deviation, are listed in Table \ref{tab:01}.

\begin{table}[h!] \centering
\begin{tabular}[t]{r c c}
 & mean value & standard deviation \\ \hline
\textbf{f\textsubscript{b1}} & 2,263 Hz & 0,013 Hz \\
\textbf{f\textsubscript{b2}} & 2,085 Hz & 0,024 Hz \\
\textbf{f\textsubscript{b3}} & 3,752 Hz & 0,019 Hz \\
\textbf{f\textsubscript{t1}} & 7,906 Hz & 0,068 Hz \\ \hline
\end{tabular}
\caption{Experimentally determined natural frequencies of the Lillebælt scale model for the first three bending modes and first torsional mode.}
\label{tab:01}
\end{table}

\subsection{Determination of the damping ratios}
In a further experiment, the damping ratios associated with the first three bending modes and the first torsional mode of the bridge model are determined. Analogous to the procedure employed for the identification of the natural frequencies, the amplitudes of the acceleration and angular velocity responses of the scale model to an external excitation are recorded. For the estimation of the damping ratios, two different approaches are pursued. First, the stochastic subspace identification (SSI) method is applied. Then the damping ratio is derived from the logarithmic decrement.

\subsubsection{Excitation}
\begin{figure}[h]
\centering
\includegraphics[width=6cm]{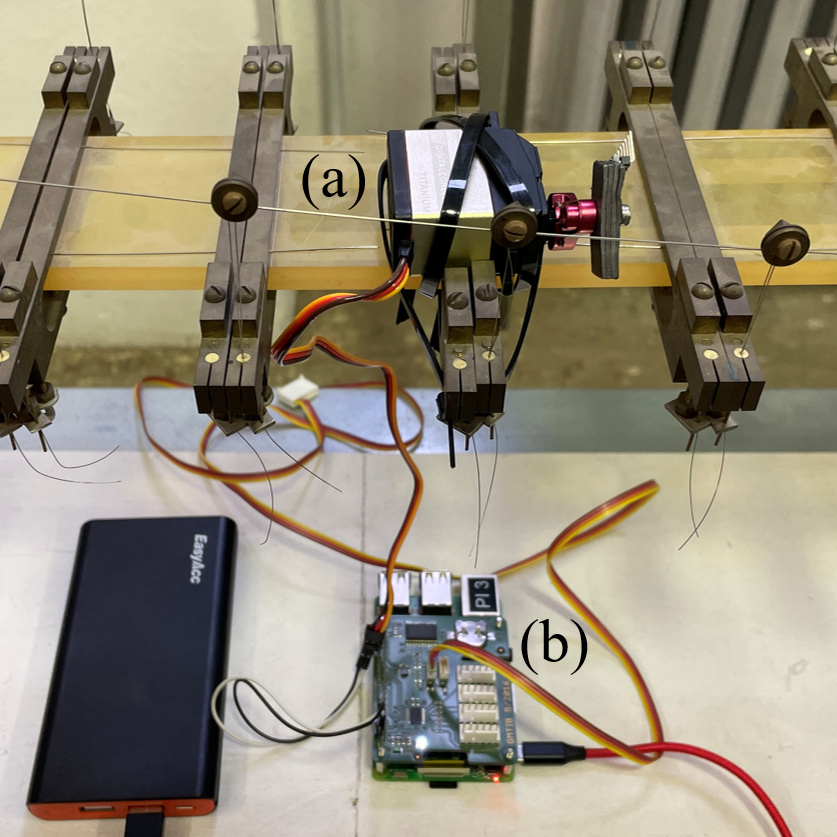}
\caption{Servomotor (a) installed on the model superstructure, connected to a Raspberry Pi (b).}
\label{fig:08}
\end{figure}

Unlike in the previous experiment, the physical model is now excited by a servomotor, which is mounted on the superstructure and controlled by a Raspberry Pi microcomputer (see Figure \ref{fig:08}). The rotational motion of an eccentric mass of $m$~=~24~g fixed to a lever arm of length $l$~=~2,9~cm induces oscillations in the bridge model. The motor is capable of rotation amplitudes up to ±30~° and provides reliable responses for oscillation frequencies up to 10~Hz \cite{Tondo.2025}.

\textbf{}
By operating the servomotor at the previously identified natural frequencies, the corresponding mode shapes can be selectively excited. The associated damping ratios are then determined from the measured accelerations in global z-direction and angular velocities around the global x-axis. To ensure that all relevant modes are excited, the servomotor is positioned at $x$~=~187~cm along the longitudinal axis of the main span. It is mounted eccentrically on the superstructure rather than on the centerline, thereby enabling the excitation of not only the vertical bending modes but also torsional modes.

\subsubsection{Sensor selection and configuration}
The same sensors employed in the previous experiment for the identification of the natural frequencies are used to determine the damping ratios (cf. Figure \ref{fig:04}). Due to the excitation with the servomotor, a single setup is sufficient in this experiment to capture all relevant natural modes of the model. The sensors are arranged at measurement points B, C and D (cf. Figure \ref{fig:05}).

In total four measurements are carried out, with the model being excited at one of the identified eigenfrequencies in each test. Each measurement lasts 120~s, during which the servomotor operates for 30~s before being switched off, allowing the system to oscillate freely until it returns to rest. For the evaluation, only the sensor data corresponding to the free vibration is considered.

\subsubsection{Stochastic subspace identification method}
\begin{figure}[h]
\centering
\includegraphics[width=8cm]{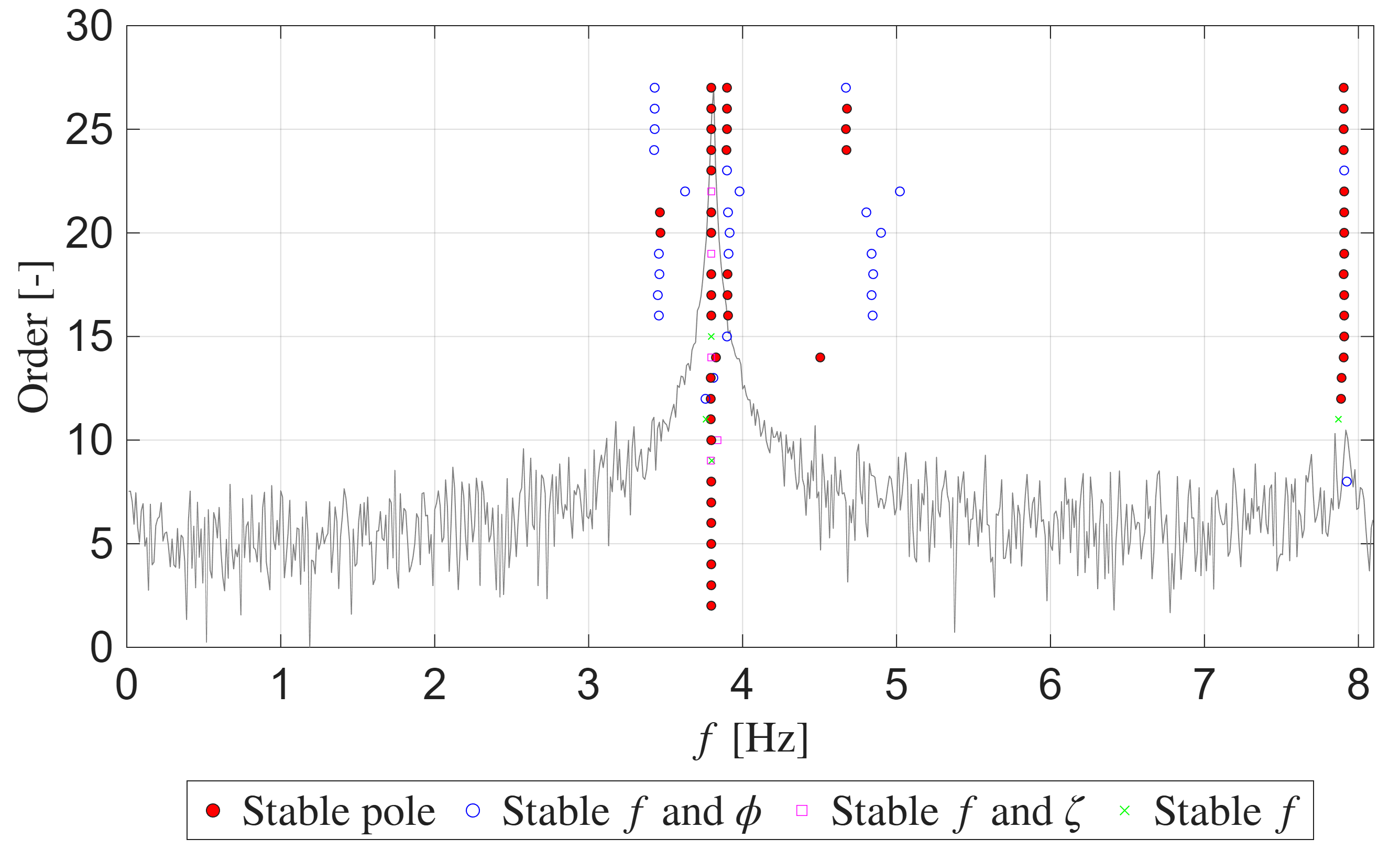}
\caption{Stabilization diagram of measured response to an excitation with the third bending eigenfrequency.}
\label{fig:09}
\end{figure}

The modal properties, including the damping ratios, are extracted from the recorded sensor data using covariance-driven stochastic subspace identification (SSI-cov) \cite{Peeters.2001}. An exemplary stabilization diagram is shown in Figure \ref{fig:09}, where pole stability is defined by a maximum error of 5~\% in the damping ratio and a maximum error of 1~\% in frequency or modal assurance criterion for increasing system orders.

\textbf{}
Determining the damping ratio using the SSI-cov method is challenging in this experiment. On the one hand, this is due to the inherently elusive nature of damping itself. On the other hand, the excitation provided by the servomotor at a constant frequency contradicts the method’s underlying assumption of excitation by white noise, which further complicates the analysis. Consequently, the damping ratios are subsequently evaluated by applying an additional method, the logarithmic decrement approach.

\subsubsection{Logarithmic decrement}
Using the logarithmic decrement, the damping ratio can be derived directly from the free decay response. The logarithmic decrement $\Lambda$ is defined as the natural logarithm of the quotient of two consecutive oscillation maxima $a\left(t_{1} \right)$ and $a\left(t_{1} + T \right)$ (see Eqn. \ref{eq:01}). If, instead of two consecutive peaks, maxima separated by an arbitrary number of $n$ periods $T$ are considered, $\Lambda$ can be calculated according to Eqn. \ref{eq:02}. The relationship between the damping ratio $\zeta$ and the logarithmic decrement is given in Eqn. \ref{eq:03} \cite{Petersen.2017}. 

\begin{align}
\Lambda &= ln \left(\frac{a(t_1)}{a(t_1+T)}\right)
\label{eq:01}
\end{align}

\begin{align}
\Lambda &= \frac{1}{n} ln \left(\frac{a(t_1)}{a(t_1+nT)}\right)
\label{eq:02}
\end{align}

\begin{align}
\zeta &= \frac{\Lambda}{2\pi}
\label{eq:03}
\end{align}
For each measurement, three free decay response histories are available, one from each sensor. The time and acceleration or angular velocity values corresponding to two peaks separated by $n$ periods are manually extracted from the curves. Using Eqns. \ref{eq:02} and \ref{eq:03}, the damping ratios are then determined for each sensor, and the mean value is subsequently calculated for each measurement.

\begin{figure}[h!]
\centering
\includegraphics[width=8cm]{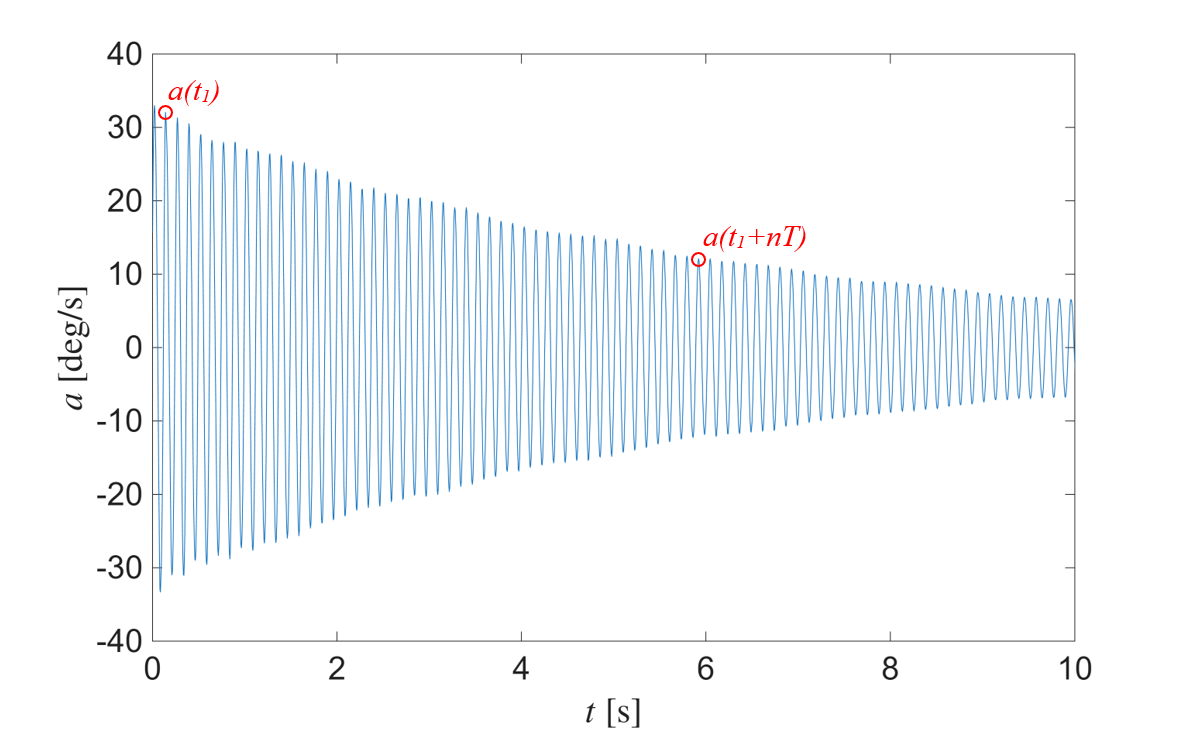}
\caption{Free decay response of the recorded angular velocity at position B under excitation at the first torsional eigenfrequency.}
\label{fig:10}
\end{figure}

Figure \ref{fig:10} shows, as an example, the free decay response of the measured angular velocities in response to an excitation at the first torsional frequency, recorded at position B.

\subsubsection{Results}
The damping ratios obtained from SSI and those calculated using the logarithmic decrement are summarized in Table \ref{tab:02}.

The values based on the logarithmic decrement represent the mean values derived from the free decay responses of the superstructure at the various sensor locations.

\begin{table}[h!] \centering
\begin{tabular}[t]{r c c}
 & SSI & log. decrement \\ \hline
\textbf{$\zeta\textsubscript{b1}$} & 0,38 \% & 0,37 \% \\
\textbf{$\zeta\textsubscript{b2}$} & 0,35 \% & 0,22 \% \\
\textbf{$\zeta\textsubscript{b3}$} & 0,24 \% & 0,19 \% \\
\textbf{$\zeta\textsubscript{t1}$} & 0,23 \% & 0,33 \% \\ \hline
\end{tabular}
\caption{Experimentally determined damping ratios of the Lillebælt scale model for the first three bending modes and first torsional mode.}
\label{tab:02}
\end{table}
\section{Relationship between scale model and full scale}
The relationship between a scale model and the full-scale structure is determined by similarity principles. These principles include geometric similarity, which requires that all dimensions are reduced by a constant scaling factor, as well as kinematic and dynamic similarity, which ensure that deformations and vibrations are accurately reproduced. To satisfy the similarity conditions, parameters such as mass distribution and stiffness must be adjusted in accordance with the scaling laws, allowing the scale model to provide representative results that can be used to describe the full-scale structure \cite{Scott.2001}. 

\subsection{Transferring results from Lillebælt model tests to full scale}
\begin{figure}[h!]
\centering
\includegraphics[width=7cm]{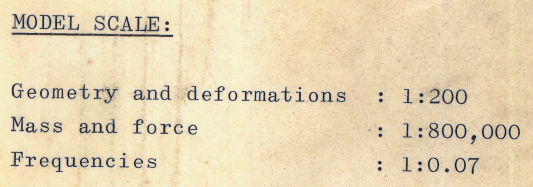}
\caption{Model scale of the Lillebælt Bridge model.}
\label{fig:11}
\end{figure}

The information sign supplied with the model, pictured in Figure \ref{fig:11}, provides details regarding the scale of the Lillebælt Bridge model.
\textbf{}

There, the frequency scale factor $F_f$~=~0,07 was determined based on the geometry scale factor $F_l$~=~200, according to Eqn. \ref{eq:04}.

\begin{align}
F_f &= \frac{1}{\sqrt{F_l}}
\label{eq:04}
\end{align}

\subsection{Full-scale frequencies}
Applying the frequency scaling factor $F_f$~=~0,07, the natural frequencies obtained from the scale model tests, listed in Table \ref{tab:01}, can be converted to full scale. The resulting values are presented in Table \ref{tab:03}, along with reference values determined during an operational modal analysis of the real Lillebælt Bridge \cite{Christensen.2019}.

\begin{table}[h!] \centering
\begin{tabular}[t]{r c c}
 & scaled values of model tests & values identified on real bridge \cite{Christensen.2019} \\ \hline
\textbf{f\textsubscript{b1}} & 0,158  Hz & 0,156 Hz \\
\textbf{f\textsubscript{b2}} & 0,146 Hz & 0,171 Hz \\
\textbf{f\textsubscript{b3}} & 0,263 Hz & 0,258 Hz \\
\textbf{f\textsubscript{t1}} & 0,553 Hz & 0,523 Hz \\ \hline
\end{tabular}
\caption{Natural frequencies of the Lillebælt bridge obtained from an operational modal analysis on the scale model and on the real bridge.}
\label{tab:03}
\end{table}
The results of the model tests correspond to a high degree with the results from the investigations on the real bridge, especially with respect to the natural frequencies of the first and third bending mode. Larger discrepancies are observed for the second bending mode and the first torsional mode. It is important to note that maintenance and repair activities on the Lillebælt Bridge may alter its dynamic properties, which can be a reason for differences between the scaled frequencies of the physical model and those observed in the real structure.

\subsubsection{Full-scale damping ratios}
The damping ratios of the scale model are expected to match those of the real structure. In the case of the Lillebælt Bridge, however, the damping ratios of the physical model, listed in Table \ref{tab:02}, show no correspondence with those determined for the real bridge (cf. \cite{Christensen.2019}). To achieve better agreement, the scale model needs to be calibrated.

\section{Conclusions}
This paper aims to enhance awareness of the continued relevance of physical models in civil engineering. Owing to their clarity and tangibility, they remain valuable engineering tools today, particularly in research and teaching.

Determining the behaviour of a structure typically requires investigations on the real construction, which are often associated with considerable effort and cost. Due to their easier accessibility, physical models offer an ideal framework to learn methods for determining system properties such as deflections, natural frequencies, mode shapes or damping ratios, or to test novel approaches in the system identification.

When a physical model is properly scaled and calibrated, the results obtained from model tests can be transferred to full scale. The scaled results can then serve directly as reference data for calibrating numerical models of the real structure through model updating, thereby eliminating inaccuracies in the digital representation and enhancing the reliability of numerical simulations. In the present study, highly satisfactory results were achieved for the natural frequencies of the Lillebælt scale model, which could be reliably converted to full scale using the prescribed frequency scaling. With regard to damping ratios, however, a prior calibration of the physical model is required before the test results can be considered as reliable for representing the full-scale structure.

\bibliographystyle{unsrt}  
\bibliography{References_IABSE2026}

\end{document}